\documentclass[aps,prb,a4paper,showpacs,final,twocolumn]{revtex4-1} 

\usepackage{hyperref}
\hypersetup{
	pdfauthor={T.~V.~Dinh, A.~Valavanis, L.~J.~M.~Lever, Z.~Ikoni\'c and R.~W.~Kelsall},
	pdftitle={An extended density matrix model applied to silicon-based terahertz
	quantum cascade lasers},
	pdfkeywords={{density matrix} {quantum cascade lasers} {terahertz} {coherent transport} {silicon} {germanium} {SiGe} {intersubband transitions} {interdiffusion}}
}

\usepackage{amsmath}
\usepackage{graphicx}
\usepackage{dcolumn}
\usepackage[caption=false]{subfig}
\begin{document}

\title{An extended density matrix model applied to silicon-based terahertz
quantum cascade lasers}

\author{T.~V.~Dinh}
\author{A.~Valavanis}
\email{a.valavanis@leeds.ac.uk}
\author{L. J. M. Lever}
\author{Z. Ikoni\'{c}}
\author{R. W. Kelsall}
\affiliation{Institute of Microwaves and Photonics, School of Electronic 
and Electrical Engineering, University of Leeds, Leeds LS2 9JT, 
United Kingdom}



\begin{abstract}
Silicon-based terahertz quantum cascade lasers (QCLs) offer potential 
advantages over existing III--V devices.  Although coherent electron transport
effects are known to be important in QCLs, they have never been 
considered in Si-based device designs.  We describe a density matrix 
transport model that is designed to be more general than those in previous
studies and to require less \emph{a priori} knowlege of electronic 
bandstructure, allowing its use in semi-automated design procedures. The 
basis of the model includes all states involved in interperiod transport, and 
our steady-state solution extends beyond the rotating-wave approximation by 
including DC and counter-propagating terms.
We simulate the potential performance of bound-to-continuum Ge/SiGe QCLs and
find that devices with 4--5-nm-thick barriers give the highest simulated optical
gain. We also examine the effects of interdiffusion between Ge and SiGe 
layers; we show that if it is taken into account in the design, 
interdiffusion lengths of up to 1.5\,nm do not significantly affect the simulated 
device performance.
\end{abstract}

\maketitle

\section{Introduction}
Terahertz quantum cascade lasers (THz QCLs) are compact, coherent radiation
sources in which electrons are transported through a periodic semiconductor 
heterostructure, with a radiative transition in each 
period.\cite{NatureKohler2002}  Silicon-based THz QCLs may offer
significant advantages over the existing III--V devices including the
absence of Reststrahlen absorption, which may allow emission at 
frequencies above the current limit of 4.9\,THz in GaAs/AlGaAs 
QCLs.\cite{APLLee2006}  III--V devices require cryogenic cooling (currently to 
$< 200$\,K\cite{OptExpFathololoumi2012}) but the high thermal conductivity of Si and the 
absence of polar LO-phonon interactions could potentially overcome this
limitation.  Additionally, there is the prospect of leveraging the mature Si 
process technology, to create 
affordable integrated electrically-driven semiconductor THz lasers. 
Although mid-infrared\cite{ScienceDehlinger2000} and THz\cite{APLLynch2002} 
electroluminescence from $p$-type SiGe/Si quantum cascade structures has been
achieved, lasing has not yet been demonstrated.  In recent years, attention 
has switched to $n$-type devices owing to the simpler electron dispersion.  We 
recently showed theoretically\cite{PRBValavanis2011} that the low effective 
mass and large usable conduction band offsets of the $L$-valleys in Ge/GeSi 
heterostructures\cite{APLLever2009,APLDriscoll2006} could allow significantly 
higher gain, operating temperature and emission frequency range than equivalent
Si/SiGe devices.\cite{IJHSESKelsall2003, PRBValavanis2008_2, APLLever2008}

Several theoretical studies of Si-based QCLs 
exist,\cite{PRBValavanis2011, APLLever2009, APLDriscoll2006, JAPDriscoll2007, 
PRBValavanis2008_2, APLLever2008, APLSun2007} but
none have accounted for coherent transport effects
(\emph{i.e.}, quantum tunneling and interactions with optical fields).
Although semiclassical scattering-transport models can give good agreement 
with experimental results,\cite{NatureKohler2002, JAPJovanovic2006} they neglect 
tunneling across barriers, and can predict unrealistically large 
spikes in current density and gain when electrons scatter between
spatially-extended subbands.\cite{JAPCallebaut2005}  By contrast, simplified 
density matrix (DM) models\cite{JAPCallebaut2005,APLKhurgin2009,PRBKumar2009,
	Dupont2010,Terazzi2010,Talukder2011a,Talukder2011b}
account for tunneling, in addition to scattering, and
include the effect of the optical field on the electron dynamics.  Additionally,
DM models are much faster and less computationally demanding than full quantum 
mechanical simulations based on Green's functions.\cite{Wacker2002,Kubis2008,
Kubis2009a,Yasuda2009,Savic2007,Weber2009}
To our best knowledge, all existing DM models of QCLs consider 
coherence between a reduced set of basis states including a single ``injector'' 
state (adjacent to a thick tunneling barrier) and a number of states in the next
period of the QCL.  This requires the manual selection of the injector 
state prior to simulating the device, and omits tunneling through the injection
barrier from other states.  This approach is not well-suited to semi-automated
design procedures,\cite{Ko2010,PRBValavanis2011} and can be problematic in 
bound--to--continuum (BTC) designs, in which multiple states may contribute to
the tunneling current.  Furthermore, these simplified models 
neglect coherences between the injector state and other states in the same 
period.
In this work, we present a generalized DM model that reduces the requirement 
for \emph{a priori} knowledge of the electronic bandstructure by including all
states involved in interperiod transport, and by including contributions from 
both the optical field and the external DC bias in all density terms in our
steady-state solution.

Although the thickness of the injection barrier is of great importance in
III--V QCls,\cite{Elec_Lett_Luo_2007} it has never been investigated in
Ge/GeSi devices.  We therefore use our DM model in conjunction with a 
semi-automated design algorithm\cite{Ko2010} to investigate the influence of 
injection barrier thickness upon the simulated device performance.
Ge/GeSi interfaces also exhibit significantly more elemental 
interdiffusion than the GaAs/AlGaAs epitaxy used in existing QCLs.  We
use our DM model to assess the effect of interdiffusion on the simulated
population inversion and gain and we compensate for the gain-reduction through
design optimization.

\section{Theoretical model}
\subsection{Bandstructure calculation}
The optically-active region of a QCL consists of a periodic semiconductor
heterostructure.  In the DM model, the ``injection barrier'' that separates
periods of the structure is assumed to be sufficiently thick that
interperiod transport is limited to quantum tunneling only.  QCL periods may be 
further subdivided into a number of modules that are separated by thick 
tunneling barriers.

We used a model-solid approximation to determine the conduction band profiles 
for the QCL structures in our simulations\cite{PRBVanDeWalle1986} with the
bandstructure parameters listed in Ref.~\onlinecite{PRBValavanis2011}. A 
self-consistent one-dimensional 
Poisson--Schr\"odinger solver was then used to locate the quasibound states 
within each module of the device, giving a total of $N$ subbands within
each period.  Localized wavefunctions $\psi_i(z)$ with energies $E_i$ were 
obtained, according to a `tight-binding' scheme, by embedding each module of the
structure between a pair of thick barriers, such that the amplitude of the 
evanescent waves decays to zero before reaching the edge of the simulation 
domain.  Here, the subscript $i\in[1\ldots N]$ denotes
the index of each state in the period, in ascending order of energy.  
Intervalley
mixing has previously been shown to yield negligibly small energy splitting
for intersubband transitions in Si-based heterostructures longer than a few 
nanometers,\cite{PRBValavanis2007, PRBVirgilio2009} and we therefore omit the
effect from our model.

As the QCL is a periodic heterostructure in an electric field, the states 
localized in other periods of the device are obtained by simple translations 
of these solutions in energy and space.  The $p^{\text{th}}$ downstream period
of the cascade therefore has states with energy $E_i^{(p)} = E_i - eFLp$, where 
$F$ is the applied electric field, $e$ is the unit charge, and $L$ is the
length of a period.  The corresponding wavefunctions are given by 
$\psi_i^{(p)}(z) = \psi_i(z - Lp)$.
The wavefunction for an electron in the system is expressed in this basis
as
\begin{equation}
\Psi(z) = \sum\limits_{i=1}^N \sum\limits_{p=0}^P c_i^{(p)} \psi_i^{(p)}(z),
\end{equation}
where $c_i^{(p)}$ is the weighting of each basis state and $P$ is the number
of periods contained in the model.\cite{JAPCallebaut2005}

\subsection{Density matrix formulation}
Density matrix calculations rely on the selection of suitable basis 
states for coherent transport through the QCL.  Conventional approaches use $N$ 
basis states to model transport across a single injection barrier.
A single ``injector'' subband is selected from the period upstream of the 
barrier.  
The remaining $N-1$ states are then selected from the period after the barrier, 
and the coherences between these states and the injector describe interperiod
tunneling transport. Some approaches simplify the calculation further by 
selecting only a subset of the states from the period after the barrier.
The manual selection of an ``injector'' subband requires \emph{a priori} 
knowledge of the electronic bandstructure, and is therefore incompatible with
semi-automated design optimization techniques.  Even with \emph{a priori}
knowledge, the selection of an injector state may not be obvious, particularly 
when the QCL is biased 
well away from subband alignment; indeed, multiple channels for interperiod
transport may exist.
Also, this limited set of basis states does not include the injector subband in the 
second period.  Although the
effect on relaxation rates is included implicitly in the simulation, the 
calculation still does not account for any coherent interactions with
this subband.
Its role in intraperiod tunneling transport and its contribution to
resonant optical transitions are therefore omitted.

Here we describe a different DM model that uses 
all the subbands localized in three adjacent periods of the QCL as 
its basis.  This method allows coherences to be calculated between all pairs of 
states in the central period, and allows interperiod tunneling (both in and 
out of the central period) to be determined without the need to select an
injector subband manually.
The resulting $3N \times 3N$ density matrix may be expressed in block form as
\begin{equation}
\mathsf\rho
=
\begin{pmatrix}
\mathsf\rho_{\text{CC}} & \mathsf\rho_{\text{CU}} & \mathsf\rho_{\text{CD}} \\
\mathsf\rho_{\text{UC}} & \mathsf\rho_{\text{UU}} & \mathsf\rho_{\text{UD}} \\
\mathsf\rho_{\text{DC}} & \mathsf\rho_{\text{DU}} & \mathsf\rho_{\text{DD}}
\end{pmatrix},
\label{eq:rho3}
\end{equation}
where the subscripts $U$, $C$ and $D$ denote the upstream, center and downstream
periods of the structure respectively.  Each of these $N \times N$ blocks 
consists of density terms for pairs of states in the periods denoted by the 
block subscripts, for example,
\begin{equation}
\mathsf\rho_{\text{CC}}=
\begin{pmatrix}
\rho_{1,1} & \ldots & \rho_{1,N} \\
\vdots & \ddots & \vdots \\
\rho_{N,1} & \ldots & \rho_{N,N}
\end{pmatrix}.
\end{equation}
The density terms are unknown values to be solved, which represent the 
ensemble average of the weightings for the basis states 
$\rho_{i,j} = \overline{c^*_ic_j}$.

The Hamiltonian for the three-period system is 
written in block form as
\begin{equation}
\mathsf{H}
=
\begin{pmatrix}
\mathsf H_{\text{CC}} & \mathsf \Omega_{\text{CU}} & \mathsf \Omega_{\text{CD}} \\
\mathsf \Omega_{\text{UC}} & \mathsf H_{\text{UU}} & \mathsf \Omega_{\text{UD}} \\
\mathsf \Omega_{\text{DC}} & \mathsf\Omega_{\text{DU}} & \mathsf H_{\text{DD}}
\end{pmatrix}.
\label{eq:ham3}
\end{equation}
Here, the off-diagonal blocks contain the Rabi frequency terms for coupling 
between states in different periods of the QCL, which were calculated according
to the scheme in Ref.~\onlinecite{Yariv1985}. The diagonal blocks such as 
$\mathsf H_{\text{CC}}$ denote the Hamiltonian matrix for a single period of the
structure.  The elements of these single-period Hamiltonians are either
the basis state energies (for diagonal elements), the Rabi frequency terms 
(for pairs of states in different modules) or the optical-coupling terms
$H_{i,j} = z_{i,j} A_\text{inc}$ for radiative transitions between states in 
response to incident light with the electric field
$A_\text{inc} = A_0\exp(\mathrm{i} \omega_0t)$.  Here, the dipole matrix
element terms are given by $z_{i,j} = \left\langle\psi_i|z|\psi_j\right\rangle$.

The time evolution of the density matrix is expressed by the Liouville equation
\begin{equation}
\frac{\partial\mathsf\rho
}{\partial t}=-\frac{\mathrm i}{\hbar}
\left[\mathsf H,\mathsf\rho\right] - \left(\frac{\partial\mathsf\rho
}{\partial t}\right)_{\text{relax}},
\label{eq:liouville}
\end{equation}
where the last term is the matrix containing all relaxation and dephasing times.

Although the Liouville equation for our three-period model contains a total of 
$9N^2$ differential equations (compared with $N^2$ for a single-period model),
the translational invariance of the system simplifies the calculation 
considerably.  Firstly, the density terms may be translated between blocks of
the matrix, such that $\mathsf\rho_{\text{UU}}=\mathsf\rho_{\text{CC}} = 
\mathsf\rho_{\text{DD}}$, 
$\mathsf\rho_{\text{CD}}=\mathsf\rho_{\text{UC}}$, and 
$\mathsf\rho_{\text{DC}}=\mathsf\rho_{\text{CU}}$.  
Similar translations may also be
applied to the Rabi frequencies, dipole matrix elements, and relaxation times.
The Liouville equation (\ref{eq:liouville}) can therefore be reduced to $3N^2$
independent differential equations, coming from the three blocks in the upper 
left corner of Eq.~(\ref{eq:rho3}) and (\ref{eq:ham3}).  Secondly,
we apply the nearest-neighbor approximation, 
so that there is no coupling or scattering between states
spaced by more than by one period. Therefore, the coupling constants between the 
first and third periods $\mathsf \Omega_{\text{DU}}$ are set to zero in 
Eq.~(\ref{eq:ham3}), and the same applies to the coherence and relaxation terms.
These simplifications reduce the computational complexity of our model to a
level comparable with the single-period model.

The relaxation matrix can be written symbolically in the form
\begin{equation}
\tau^{-1}=\left[
\begin{array}{ccc}
\left(1/\tau_{\text{CC}},1/\tau_{\|\,\text{CC}}\right) & \left(1/\tau_{\|\,\text{CU}}\right) & - \\
\left(1/\tau_{\|\,\text{UC}}\right) & - &  - \\
- & - & -
\end{array} \right]
\label{eq:relax}
\end{equation}
where the dashes indicate that the corresponding $N\times N$ block
is irrelevant owing to translational invariance. The relaxation 
terms in Eq.~(\ref{eq:relax}) determine both the linewidths of optical
transitions and tunneling lifetimes.  They contain
contributions from the state lifetimes $\tau_{i,j}$, as well as `pure dephasing' 
times $\tau_{\|\,i,j}$ for intraperiod transitions.\cite{JAPCallebaut2005}  The 
off-diagonal blocks in
Eq.~(\ref{eq:relax}) (denoted $\tau_{\|\,\text{UC}}$, etc) describe the pure
dephasing for interperiod interactions.

In this work, subband relaxation rates and tunneling dephasing rates were 
calculated by accounting for all relevant scattering
processes:\cite{PRBValavanis2011} acoustic and optical phonon scattering, 
intervalley scattering, ionized impurity and interface roughness scattering.
The pure 
dephasing contributions are not straightforward to calculate, and a 
number of different schemes for their estimation have been 
proposed.\cite{APLKhurgin2009,Talukder2011a,
Talukder2011b}  
In the case of tunneling transitions, the prescription from 
Refs.~\onlinecite{Talukder2011a,Talukder2011b} was employed.
In case of optical transitions, however, we have simply set 
the linewidths to 2\,meV (as is typical for GaAs THz QCL 
structures).\cite{NatureKohler2002}

\subsection{Steady-state solution}
The harmonic balance method is a convenient approach for determining a 
steady-state solution of the Liouville equation.  Here, a simple steady-state 
functional form is assumed for each element of the density matrix depending on
the states involved.  Under the rotating-wave approximation (RWA), 
each element is assumed to contain only a single frequency harmonic. The 
diagonal elements of the density matrix $\rho_{i,i}$ give the state populations,
and are assumed to be constant-valued.  The off-diagonal terms $\rho_{i,j}$ 
describe coherences between states.  In the RWA approach, the steady-state
forms of these terms are selected depending on whether they are optically-active
or not.  It is assumed that pairs of states within the same period with an
energy separation
$E_{ij} \sim \hbar\omega_0\pm\Gamma$ will interact strongly with the driving 
optical field (including a transition linewidth $\Gamma$). 
The density matrix terms for optical transitions are assumed to be proportional 
to $\exp(\mathrm{i}\omega_0t)$ for $E_i<E_j$ or $\exp(-\mathrm{i}\omega_0t)$ for
$E_i > E_j$.  All other off-diagonal elements, (\emph{i.e.}~for pairs of states 
with
small energy separations, or in different modules of the structure) are assumed 
to represent tunneling transport and are assumed to be constant valued. With the
single dominant harmonic assumed for each $\rho_{i,j}$, the Liouville equation
becomes a system of $3N^2$ ordinary linear equations. 
One of the equations is then replaced by the particle conservation law,
$\operatorname{Tr}(\rho)=1$, and the system becomes inhomogeneous, with a
unique solution.

Although the RWA is useful for rapidly calculating the density matrix for a
known system, it is incompatible with semi-automated design tools where the
state energies are not known \emph{a priori}.  The RWA also potentially omits 
multi-frequency effects on density matrix terms.  In this work, therefore, we 
use an enhanced ``non-RWA'' solution method, that allows three harmonic terms to
be included in the density matrix elements such that
\begin{equation}
\rho_{i,j} = \rho^+_{i,j}\exp(\mathrm{i}\omega_0t) + \rho^{\text{DC}}_{i,j} + 
\rho^-_{i,j}\exp(-\mathrm{i}\omega_0t),
\end{equation}
where $\rho^{\pm}_{i,j}$, and $\rho^{\text{DC}}_{i,j}$ are unknown amplitudes
for each of the harmonic components.  As the Liouville equation is linear, each
harmonic term may be treated independently, resulting in a system of up to 
$9N^2$ linear equations (if every term contains all three 
harmonics), which still presents relatively low computational complexity. 
Similarly, in the light-field interaction terms in the Hamiltonian both the 
$\exp(\pm\mathrm i\omega_0 t)$ components are retained.  

Previous studies indicate that the counter-propagating non-RWA terms may 
measurably affect the gain of lasers or the intensity-induced shift
of resonance frequency in light-matter interactions, but are not very
significant in near-resonant laser operation.\cite{Casperson1992,Zubairy2009,
Gil2011}  Indeed, in all the cases considered in this work, we find that only 
one of the three harmonic amplitudes is significant in our simulations, 
which validates the use of the RWA when \emph{a priori} knowledge of the state 
energies exists.

The calculation can include thermal self-consistency (energy balance) by 
allowing the electron temperature $T_e$ to be variable and requiring the total 
energy exchange between the electron gas and the lattice to equal 
zero.\cite{JAPJovanovic2006} In this 
work we did not include thermal balance within the density matrix equations, and
have instead used the electron temperature (assumed common to all subbands) as 
delivered by the rate equations model.\cite{PRBValavanis2011}
This is expected to be a reasonable approximation, since the tunneling 
contribution\cite{Beji2011} is not the major heat generating process in QCLs.

\subsection{Output parameters}
The current is calculated from the density matrix as 
$j=\operatorname{Tr}(\mathsf{\rho J})$,
where 
\begin{equation}
 \mathsf{J}=e\frac{\mathrm{i}}{\hbar}\left[\mathsf{H},\mathsf{z}\right]. 
\end{equation}
The current has
both a time-independent (DC) component and a harmonic (AC) component that 
is induced by the optical field. The latter is 
used to find the complex permittivity
$\tilde\epsilon$ of the electron gas of the active medium, from
\begin{equation}
j_{\text{AC}}=\tilde\epsilon \frac{\mathrm{d}}{\mathrm{d}t}A_{\text{inc}},
\end{equation}
and then the gain
from $g\approx\omega_0\cdot\text{Imag}(\tilde\epsilon)/n_rc$, where $n_r$ is
the refractive index and $c$ is the speed of light in vacuum. Using very small 
values of $A_{\text{inc}}$ gives the small signal gain of the QCL active region
(\emph{i.e.}~in the absence of gain saturation).

\section{Simulation results}
\begin{figure}
\includegraphics*[width=8.6cm]{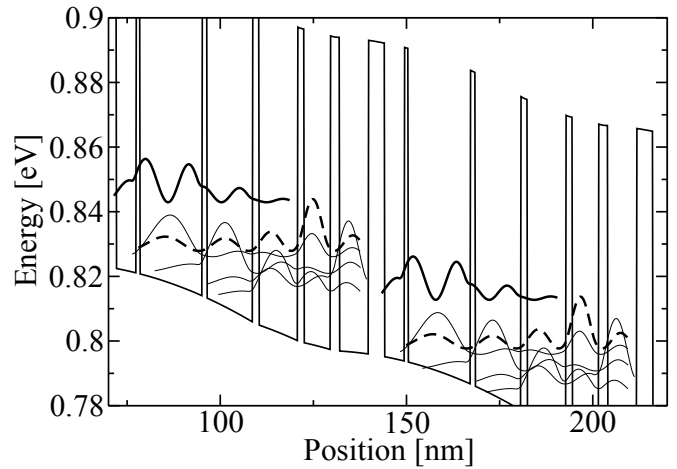}
\caption{\label{fig:BTC-6well-no-diff} 
Two periods of the bandstructure of the six-well BTC QCL design which serves as 
a template for our investigations.  The thicknesses of the layers are as
follows (starting from the injection barrier): \textbf{4.3}/5.0/\textbf{1.2}/14.3/\textbf{1.3}/12.9/\underline{\textbf{1.6}}/7.9/\textbf{\underline{1.8}}/7.1/\textbf{\underline{2.5}}/7.8
where bold text denotes the Si$_{0.15}$Ge$_{0.85}$ barriers and regular text 
denotes Ge wells.  The underlined layers are doped to provide a sheet doping 
density of $8\times10^{10}$\,cm$^{-2}$.  The upper and lower laser level in
each period of the structure are shown as bold, and dashed-bold lines
respectively.
}
\end{figure}
We have recently reported the use of a semi-automated design optimization 
process to show that THz QCLs in the (001) Ge/GeSi material configuration yield
substantially higher simulated gain than equivalent (111) and (001) 
Si/SiGe designs.\cite{PRBValavanis2011}
This method affords a systematic comparison between different device
design schemes and/or materials systems.  In this work, we have used our 
optimization algorithm, in conjunction with the DM model described above, to 
identify viable designs for bound--to--continuum (BTC) Ge/SiGe QCLs, while
accounting for coherent transport effects.  Previous DM models of III--V BTC 
QCLs have 
reproduced experimental results by including only the upper laser level (ULL) 
and a subset of miniband states in their basis set, of which one is designated 
as the injector.\cite{APLKhurgin2009}  However, in practice multiple subbands
may contribute significantly to interperiod tunneling in BTC QCLs.
Our extended non-RWA DM model avoids the need for \emph{a priori}
selection of the optically-coupled transition and the injector subband, allowing
us to use the semi-automated design approach described in 
Ref.~\onlinecite{Ko2010}.  Furthermore, our model explicitly includes all 
possible interperiod tunneling pathways.

Figure~\ref{fig:BTC-6well-no-diff} shows two periods of the bandstructure of a
4\,THz six-well BTC QCL design which serves as a template for our investigation
of device performance. For our DM simulations, each period of the active-region 
structure was modeled as a single module.  All the devices simulated in this
work were derived from this template by systematically adjusting a single
element of the structure (\emph{i.e.}~either the injection barrier thickness or 
the 
interdiffusion length).  In all cases, a lattice temperature of 4\,K was used. 
The electron temperature was fixed at a value of 100\,K, which was obtained from
a semiclassical simulation of the design template.

As Si and Ge are not lattice matched, the proposed QCL structures must be grown
on a relaxed buffer or ``virtual substrate''. Such substrates can be achieved
either by growth of a linearly graded alloy layer, from pure silicon up to the
desired buffer composition,\cite{APLRossner2004,PRBBonfanti2008} or by growth of
a Ge seed layer
directly on a silicon wafer, followed by reverse linear grading from pure Ge
down to the buffer composition.\cite{Shah2008,JAPShah2010} The relaxed buffer 
composition is
calculated to achieve ``strain symmetrization'' throughout the QCL 
structure,\cite{Harrison2005} whereby the compressive stress in the Ge 
wells is balanced by the tensile stress in the barriers, yielding zero net 
stress over each period.  For all the cases presented below, the optimum buffer 
composition was found to be Ge$_{0.97}$Si$_{0.03}$.

\subsection{Injection barrier thickness}
\begin{figure*}
	\subfloat[][]{
		\includegraphics*[width=0.45\textwidth]{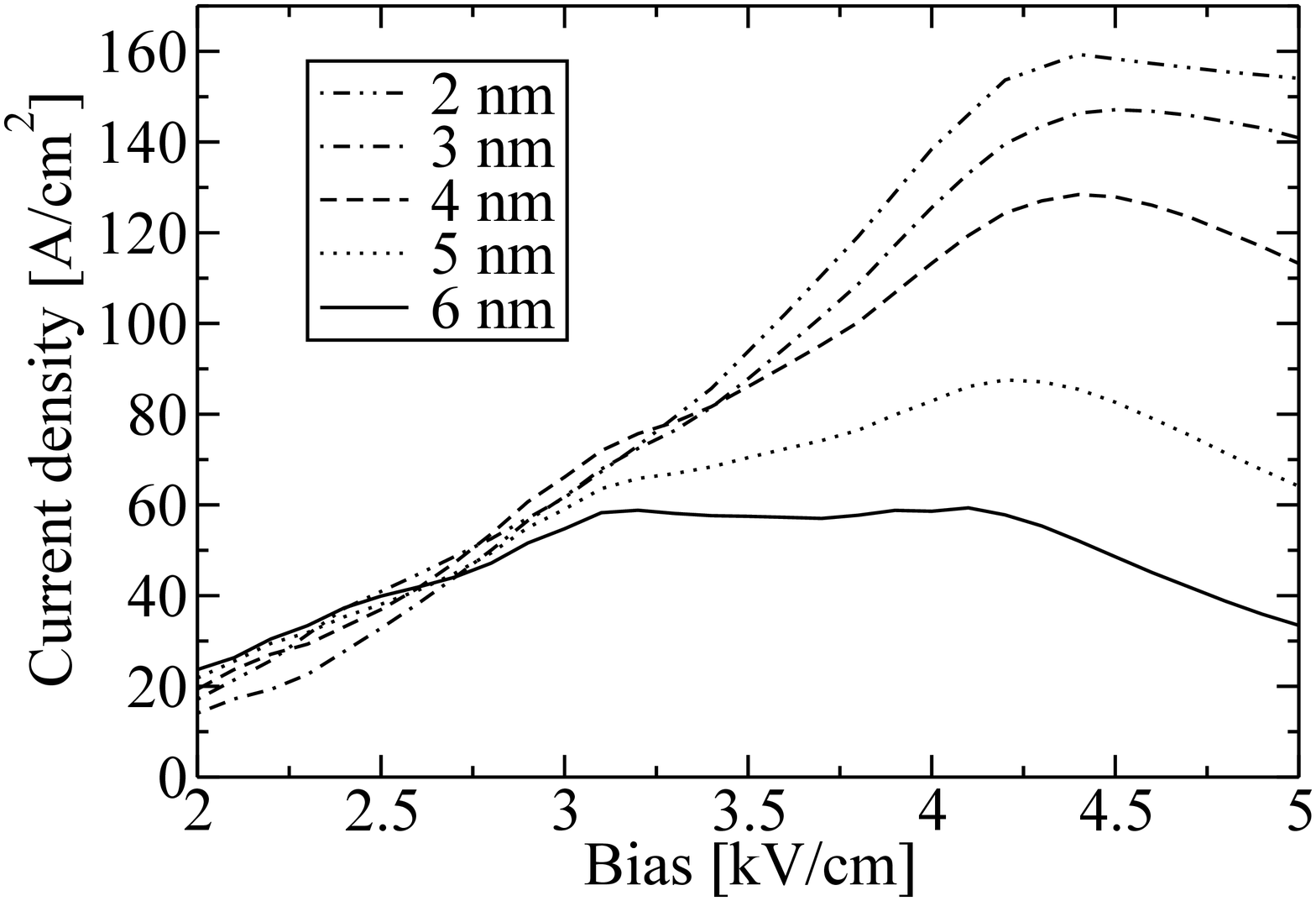}
		\label{subfig:BTC-6wells-barrier-current}
	}
	\subfloat[][]{
		\includegraphics*[width=0.45\textwidth]{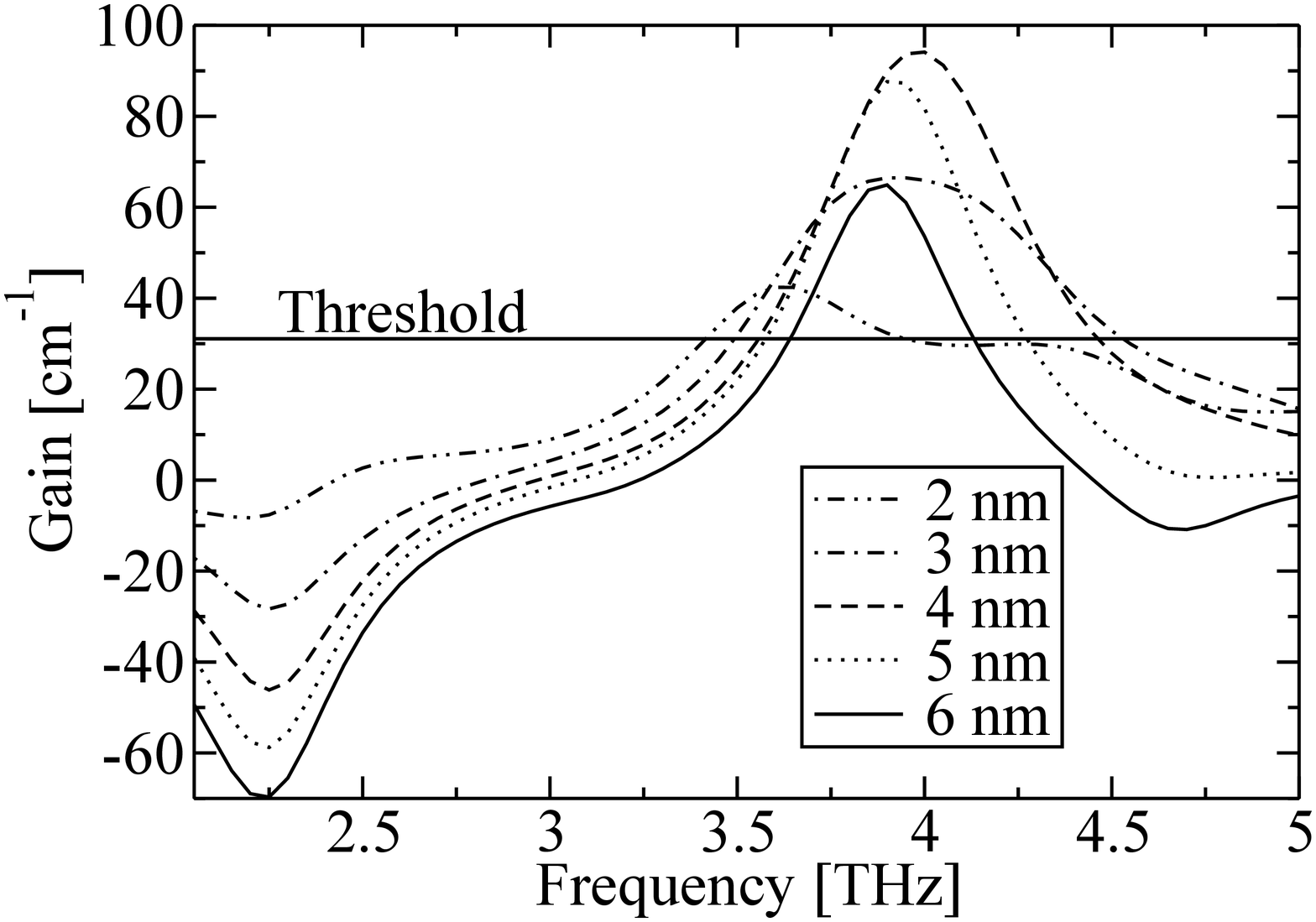}
		\label{subfig:BTC-6wells-barrier-gain}
	}
	\caption[Simulated current density and gain spectra of QCLs with
varying injection barrier thickness]{\label{fig:BTC-6wells-barrier-performance} 
	\subref{subfig:BTC-6wells-barrier-current} Simulated current density and
	\subref{subfig:BTC-6wells-barrier-gain} gain spectra of optimized 
	BTC QCLs for different injection barrier thicknesses.  An indicative
        figure for the lasing threshold is included for reference, based on 
        calculations of the waveguide losses in 
	Ref.~\onlinecite{Thesis_Valavanis2009}.}
\end{figure*}
The thickness of the injection barrier is known to be an important parameter in
III--V THz QCLs as it can significantly affect the performance of 
devices.\cite{Elec_Lett_Luo_2007}
If the injection barrier is too thin, selectivity of injection into the upper 
laser level is poor; if it is too thick, the tunneling rate through the barrier
is small and efficient injection cannot be achieved.  Semiclassical 
rate-equation models of charge injection in THz QCLs lack sensitivity to the 
injection barrier thickness, and a model that accounts for coherent effects is
therefore required.\cite{JAPCallebaut2005}
Here, we systematically vary the thickness of the injection barrier in the 
design template, and use our DM simulation to determine the gain and current
density.  We use the genetic algorithm described in Ref.~\onlinecite{Ko2010} to 
maximize the simulated gain of the device at 4\,THz by varying the thickness of 
each of the other layers in the structure, and the applied electric field.  

In all the cases considered here, the optimized layer widths (excluding the 
injection barrier) were found to be identical (to \aa{}ngstr\"om precision) to 
those of the template.  This can be understood by recalling that the
decoupled wavefunctions in the DM model are found by embedding the 
active-region module between thick barriers.
Therefore, the layer-width optimization procedure only directly affects the
single-period Hamiltonian, and has a very much weaker influence on the 
interperiod coupling (by slightly adjusting the Rabi frequencies). As such,
the results below can be seen to solely represent the effect of the
injection barrier thickness upon the device performance without including
any contribution from changes in the bandstructure within the period.

Figure~\ref{subfig:BTC-6wells-barrier-current} 
shows the simulated current density as a function of applied electric field for 
optimized devices with 2, 3, 4, 5, and 6-nm-thick injection barriers. 
Interperiod scattering between spatially-extended 
wavefunctions is avoided in the DM model, and the simulated current density 
is therefore a smoothly-varying function of bias in all cases.
The 
alignment bias for the device is approximately 4\,kV/cm for all five structures,
and we see that the current density at this bias decreases monotonically as the
thickness of the injection barrier increases.  The gain spectrum for each structure
at the alignment bias is shown in 
Fig.~\ref{subfig:BTC-6wells-barrier-gain}.
The variation in the magnitude of the peak gain with injection barrier thickness
is not monotonic, and there is an optimum thickness at around 4--5\,nm, at
which a gain of around 90\,cm$^{-1}$ is predicted.
The difference in gain spectrum between the device with 2-nm-thick injection
barriers and the other structures is caused principally by the reduction of 
injection selectivity into the ULL.  It is important to note, however, that the
``injection barrier'' in this structure is thinner than the 2.5-nm-thick barrier
at the end of the module, and the chosen subdivision of modules for tunneling 
transport in the DM model is likely to be unrealistic.

\begin{figure*}
	\subfloat[][]{
		\includegraphics*[width=0.45\textwidth]{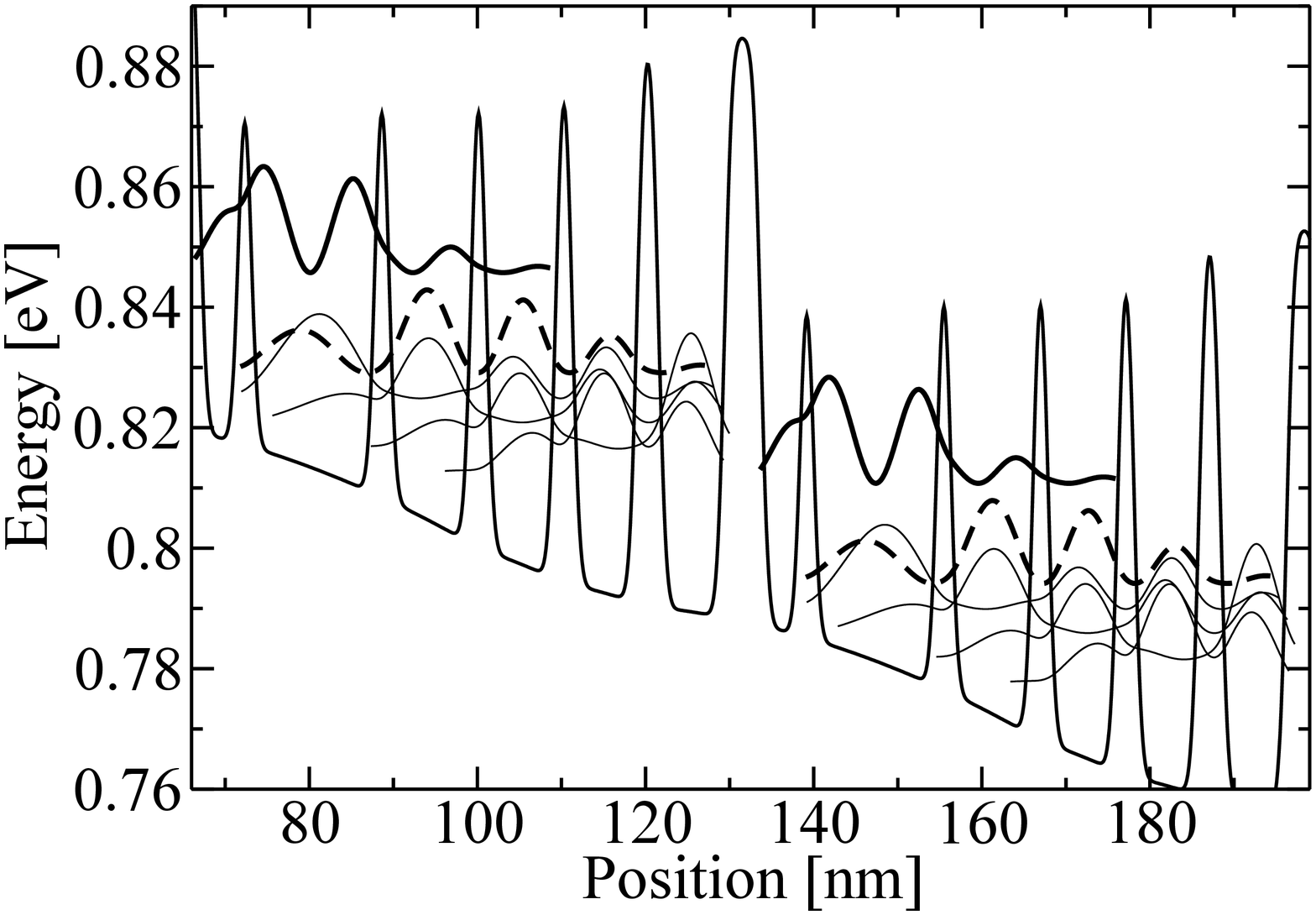}
		\label{subfig:BTC-6well-diff1}
	}
	\subfloat[][]{
		\includegraphics*[width=0.45\textwidth]{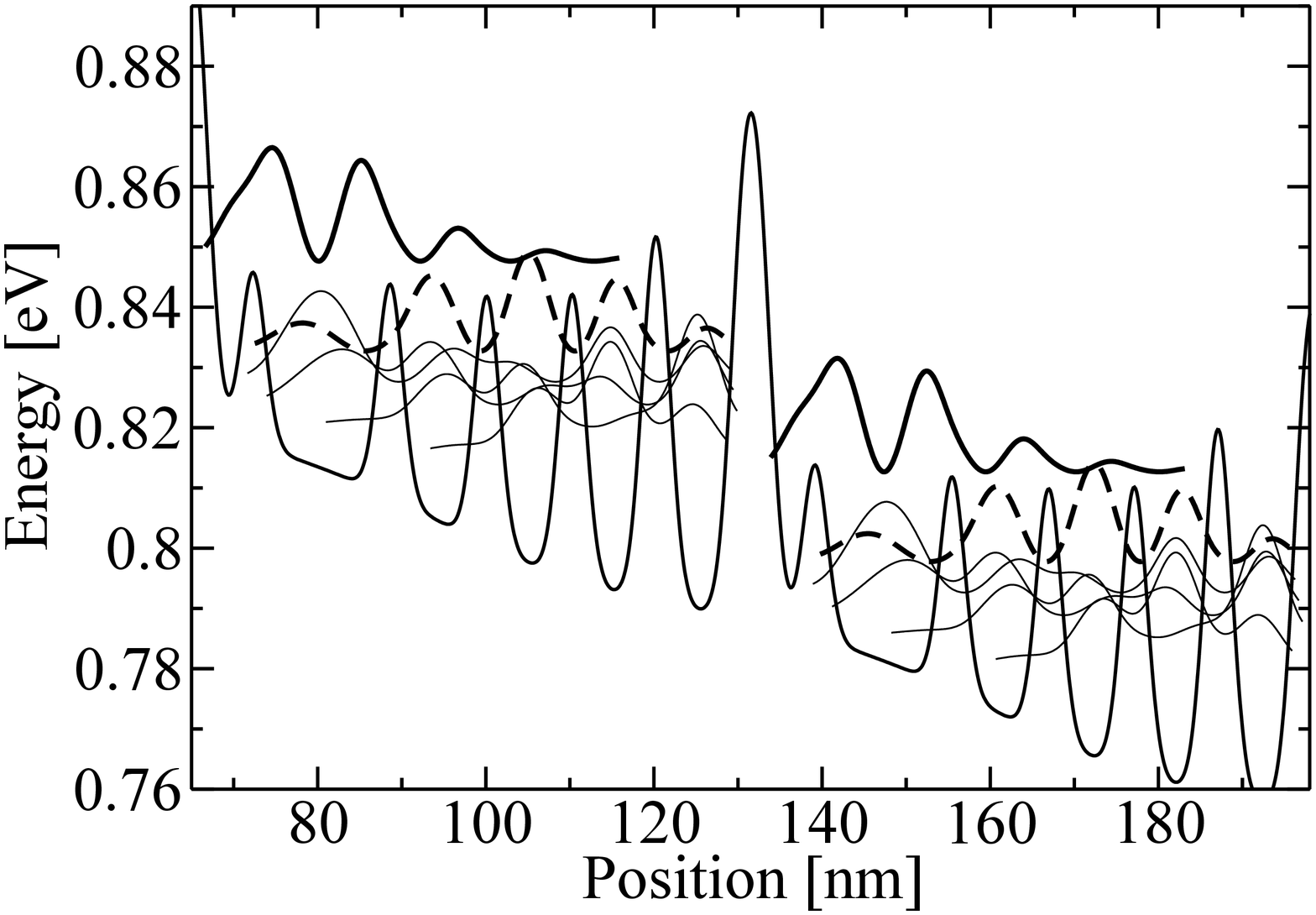}
		\label{subfig:BTC-6well-diff2}
	}
	\caption[Bandstructure of QCLs with interdiffusion]{\label{fig:BTC-6well-diff} 
Bandstructure of the six-well BTC QCLs with interdiffusion lengths of 
\subref{subfig:BTC-6well-diff1} $L_d=1$\,nm and 
\subref{subfig:BTC-6well-diff2} $L_d=2$\,nm.}
\end{figure*}

\subsection{Interdiffusion compensation}
Epitaxial Ge/SiGe heterostructures have been reported to show significant 
interdiffusion between the pure and alloy semiconductor layers, with typical 
characteristic interdiffusion lengths estimated to be of the order of 
1--2\,nm.\cite{IEEELever2010, OptLettLever2011}  This leads to significant
changes in the bandstructure and scattering lifetimes,\cite{PRBValavanis2008,
JOAValavanis2009} and can therefore degrade the performance of devices.
In this section, we investigate the impact of interdiffusion on the QCL gain
and we attempt to recover the lost performance through design optimization.

We account for the effects of interdiffusion by applying a Gaussian annealing 
model to the alloy composition profile,\cite{IEEELi1996}
such that the alloy fraction across the interface is described by a Gauss 
error function with the interdiffusion length $L_d$ as a size parameter.
Figure~\ref{fig:BTC-6well-diff} shows the bandstructure of the QCL design 
template with 1 and 2\,nm interdiffusion included.  The interdiffusion causes the 
barriers to become reduced in height and the shape of the quantum wells becomes 
distorted, with the tops being widened and the bottoms being narrowed.  For 
interdiffusion lengths of 2\,nm, the thinner barriers near the optically 
active wells are significantly reduced in height and the ULL is 
poorly confined.

\begin{figure*}
	\subfloat[][]{
		\includegraphics*[width=0.45\textwidth]{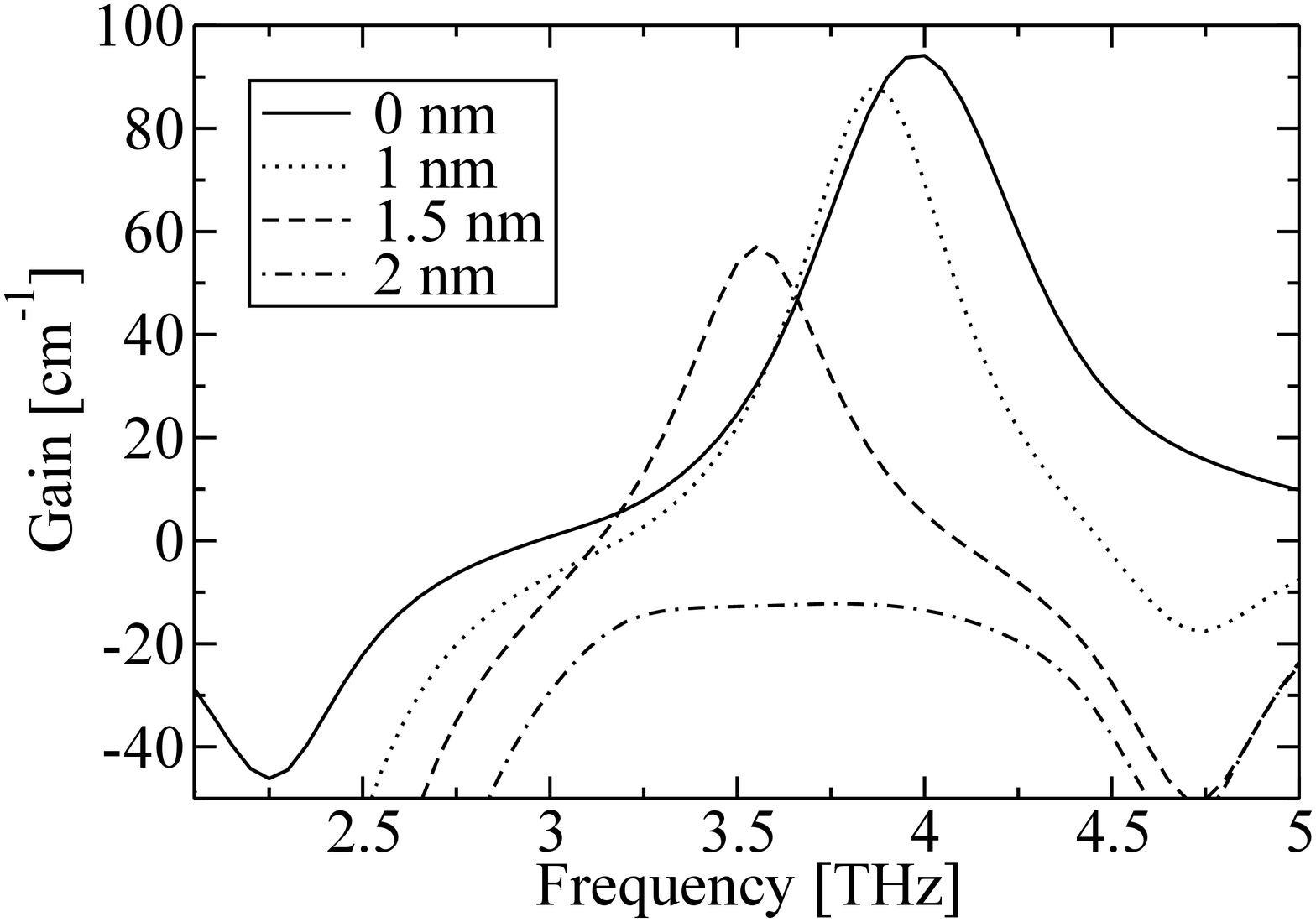}
		\label{subfig:BTC-6well-diff-no-opt}
		}
	\subfloat[][]{
	\includegraphics*[width=0.45\textwidth]{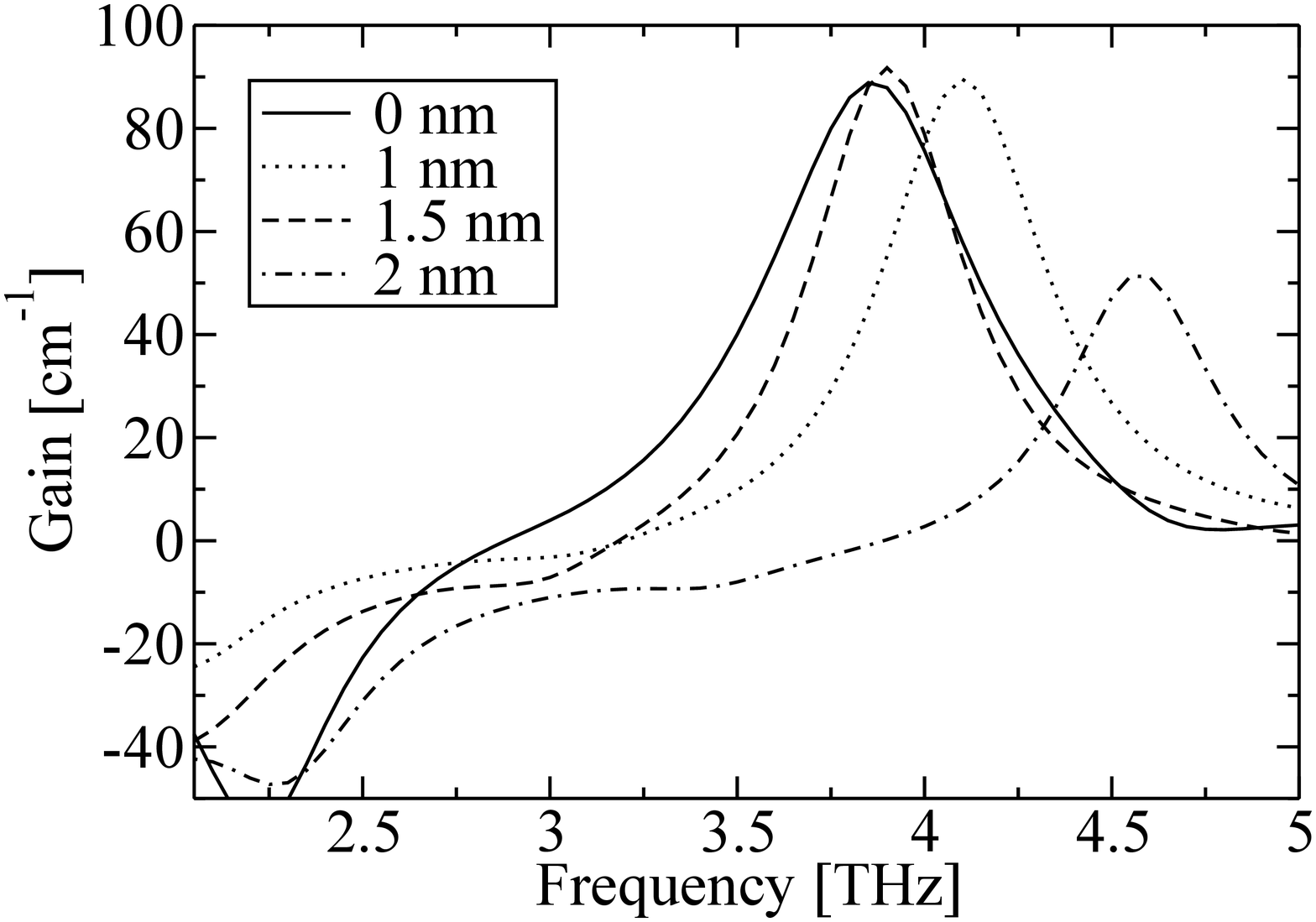}
		\label{subfig:BTC-6well-diff-opt}
	}
	\caption[Simulated gain for QCLs with interdiffusion]{\label{fig:BTC-gain-diff}
Simulated gain of the QCL for interdiffusion lengths of 1, 1.5 and 2\,nm; 
\subref{subfig:BTC-6well-diff-no-opt} without optimization and 
\subref{subfig:BTC-6well-diff-opt} where the device structure was optimized.
In each case, the legend of the plot indicates the interdiffusion length.}
\end{figure*}

Fig.~\ref{subfig:BTC-6well-diff-no-opt} shows the 
simulated gain spectra for structures with $L_d$ in the range 0--2\,nm at the
respective
operating bias for each device. Here, the interdiffusion has been included 
without subsequently optimizing the design.  There is no significant drop in the
peak gain when 1\,nm interdiffusion length is included.  However, for larger
interdiffusion lengths the simulated peak gain is reduced considerably, and 
there is no simulated gain at all for the structure with $L_d=2$\,nm.

We applied the design optimization algorithm to each of the diffuse
QCL structures, and the resulting gain spectra are shown in 
Fig.~\ref{subfig:BTC-6well-diff-opt}.  It can be seen
that the gain has been fully recovered for structures with interdiffusion 
lengths up to 1.5\,nm.  This highlights the importance of being able to 
characterize the interdiffusion length in these system: so long as this is 
known, and so long as it is 1.5\,nm or less, our results indicate that it can be
taken into account in the design process.

For interdiffusion lengths of 2\,nm or more, the gain cannot be recovered.  
There are two reasons for this: first, the ULL is no longer confined by the thin
barriers in the structure, which leads to a large spatial overlap with the 
miniband states, and hence a loss of population inversion.
Second, the interdiffusion introduces Si into the nominally pure Ge well
regions, leading to a large increase in alloy disorder scattering
rates.\cite{PRBValavanis2008}  As such, additional rapid scattering pathways are
introduced to the system, leading to rapid depopulation of the ULL.  By way of
comparison, we have previously shown that the total scattering rate within a 
Si/Ge/Si quantum well increases by 50\% at an interdiffusion length of 
1.21\,nm.\cite{JOAValavanis2009}

\section{Conclusion}
We have investigated coherent transport effects in a Si-based THz QCL
through the use of an extended density matrix model that includes in its 
basis all subbands that are involved in interperiod transport.  Our use of the 
non-rotating wave
approximation allows a steady-state solution to the Liouville equation without
\emph{a priori} knowledge of the bandstructure of the device.  In all cases, the
non-RWA solution yielded a single strongly-dominant frequency component in each
density term, indicating that it would be in good agreement with an equivalent 
RWA model.  Although we have used our generalised
non-RWA model to analyze coherent effects in Si-based QCLs, it is equally 
applicable to III--V QCL structures.

We have coupled our model with a semi-automated QCL design algorithm, and have 
shown that the optimum injection barrier thickness for Si-based BTC THz QCL 
structures is in the range
4--5\,nm, and we predict peak gain values of $\sim90$\,cm$^{-1}$ at a lattice
temperature of 4\,K.  We have also studied the effect of interdiffusion between
the Ge and GeSi layers, and found that it is possible to compensate for 
interdiffusion effects through design optimization up to a limit of 
$L_d\approx1.5$\,nm.

\begin{acknowledgments}
This work was supported by EPSRC grant EP/H02350X/1 ``Room temperature terahertz
quantum cascade lasers on silicon substrates''.
\end{acknowledgments}

\bibliography{SiGeQCL-rho}
\end{document}